\documentclass[twocolumn,prb]{revtex4}
\usepackage[utf8]{inputenc}
\usepackage{amsmath,amsfonts,amssymb,graphicx,multirow,overpic,dsfont,bm,braket}

\begin{document}
\title{
Universal quantum computation using fractal symmetry-protected cluster phases
}
\author{Trithep Devakul}
\affiliation{Department of Physics, Princeton University, Princeton 08540}
  \author{Dominic J. Williamson}
  \affiliation{Department of Physics, Yale University, New Haven, CT 06520-8120, USA}
\date{\today}

\begin{abstract}
We show that 2D fractal subsystem symmetry-protected topological phases may serve as resources for universal measurement-based quantum computation. 
This is demonstrated explicitly for two cluster models known to lie within fractal symmetry-protected topological phases, and computational universality is shown to persist throughout those phases. 
One of the models considered is simply the cluster model on the honeycomb lattice in one limit.
We discuss the importance of rigid subsystem symmetries, as opposed to global or $(\text{D}-1)$-form symmetries, in this context.
\end{abstract}

\maketitle

\section{Introduction}
\label{sec:intro}
An entangled quantum state can serve as a resource for universal quantum computation using only non-entangling (single qubit) measurements, via a scheme called measurement-based quantum computation (MBQC)~\cite{Raussendorf2001-xm,Raussendorf2003-vg,Nielsen2006-cy, Briegel2009-po, Raussendorf2012-wz,Wei2018-fi}.
A wide variety of states have been shown to be computationally useful as resources for MBQC~\cite{Van_den_Nest2006-lp,Verstraete2004-tx,Gross2007-gr,Gross2007-qp,Chen2009-bv,Wei2011-qx,Miyake2011-hm,Wei2012-zg,Wei2014-le,Wei2015-mr,Cai2010-xo,Miller2016-oz,Nautrup2015-gt,Miller2018-fe,Huang2016-lo,Wei2017-gy},
with the standard (and first) examples being the cluster states~\cite{Raussendorf2001-xm}.
The concept of computational usefulness has also been extended to \emph{phases} of matter~\cite{Doherty2009-fd,Wei2017-gy,Miyake2010-cc,Else2012-ie,Miller2015-gl,Stephen2017-cp,Raussendorf2017-gb,Darmawan2012-wg,Huang2016-lo,Williamson2015-sb}, which possess uniform computational usefulness throughout an entire phase.
In particular, this was proven generally for 1D symmetry-protected topological (SPT) phases~\cite{Else2012-ie,Miller2015-gl,Stephen2017-cp,Raussendorf2017-gb}, and is intimately related to their classification~\cite{Gu2009-ly,Chen2010-sm,Chen2011-ss,Chen2012-oa,Pollmann2012-lv,Schuch2011-gq,Chen2013-gq}.

However, MBQC is only universal in 2 or higher dimensions, as one spatial dimension must play the role of time in the quantum circuit.
In 2D, regions of computational usefulness have been shown numerically to coincide with the phase diagram of nontrival SPT phases~\cite{Darmawan2012-wg,Huang2016-lo,Wei2017-gy}, and proven to persist within small perturbations about the cluster state fixed point of the square lattice cluster model~\cite{Else2012-li}.
Recently, this same cluster model was proven to possess universal computational power everywhere within a \emph{cluster phase}, protected by rigid line-like symmetries~\cite{Raussendorf2018-nh}.
This phase is in fact a 2D \emph{subsystem} SPT~\cite{You2018-em}, which in higher dimensions are more generally related~\cite{You2018-as} to models of fracton topological order~\cite{Chamon2005-fc,Bravyi2011-fl,Yoshida2013-of,Haah2011-ny,Castelnovo2012-bq,Vijay2015-jj,Vijay2016-dr,Nandkishore2018-ee,Williamson2016-lv}.
The backbone of the proof in Ref.~\onlinecite{Raussendorf2018-nh} relies on the emergence of a \emph{symmetry-protected cellular automaton} acting on the virtual (computational) space.
Recently, \emph{fractal} subsystem SPT phases~\cite{Devakul2018-ru,Williamson2016-lv,Kubica2018-dp} have also been discovered, which are protected by fractal symmetries arising from cellular automata.  

In this paper, we show that some of the fractal SPT phases of Ref.~\onlinecite{Devakul2018-ru} constitute a computationally useful phase for universal MBQC. This provides a second provable class of such phases in 2D, after Ref.~\onlinecite{Raussendorf2018-nh}.  
The cellular automaton generating the fractal symmetries directly leads to the same symmetry-protected cellular automaton acting on the virtual space, the vital component in the proof of Ref.~\onlinecite{Raussendorf2018-nh}.
Two fractal symmetric cluster models are considered explicitly. 
We finally discuss the importance of SPTs protected by rigid (either line-like or fractal) subsystem symmetries, as opposed to \emph{higher form} SPTs~\cite{Gaiotto2015-cj,Kapustin2013-ci,Yoshida2016-zo}.

\section{Fractal Symmetric Cluster States}
Here, we first give a practical review of the fractal symmetric cluster models~\cite{Devakul2018-ru}.
The cluster state on any lattice is the unique ground state of the commuting-projector \emph{cluster Hamiltonian},
\begin{equation}
H_{\mathcal{C}} = - \sum_{s} X_s \prod_{s^\prime \in \Gamma(s)} Z_{s^\prime}  \label{eq:clusham}
\end{equation}
where $s$ denotes a site, $\Gamma(s)$ is the set of all sites connected to $s$ by an edge, and $X_s,\, Z_s$, are the Pauli matrices acting on the spin-$1/2$ degree of freedom at site $s$.
We consider \emph{symmetries} given by products of $X$ operators of the form 
$S(\{q_s\}) = \prod_{s} X_s^{q_s}$
where each $q_{s}\in \{0,1\}$ is an element of $\mathbb{F}_2$, and $[S(\{q_s\}),H_\mathcal{C}]=0$.
In the cluster models we consider here, the symmetries act on some \emph{fractal} subset of sites.
These arise naturally by considering $\{q_s\}$ as the space-time evolution of a $1$D additive cellular automaton~\cite{Yoshida2013-of,Devakul2018-ru}.


We work explicitly with two specific models, the Sierpinski cluster model (SC) and the Fibonacci cluster model (FC).  
These are defined on the square lattice with a unit cell composed of two sites, which we label as the $a$ and $b$ sublattices.  
Let us label each site by $s=(i,j,\alpha)$, where $\vec{r}(i,j) = i \vec{e}_1 + j \vec{e}_2$ give the Cartesian coordinates of the unit cell, and $\alpha\in\{a,b\}$ the specific site in the unit cell.
We take $\vec{e}_1=(1,0)$ and $\vec{e}_2=(0,-1)$, such that increasing $j$ corresponds to moving ``downwards'' in the $xy$ plane.
For convenience we denote the Pauli matrix $Z_{s} = Z_{i,j}^{(\alpha)}$, and similarly for $X$ and $Y$.  


The SC Hamiltonian is given by
\begin{eqnarray}
H_{SC} &=& -\sum_{i,j} X_{i,j}^{(a)} Z_{i,j-1}^{(b)} Z_{i,j}^{(b)}  Z_{i+1,j}^{(b)} \nonumber \\
&&-\sum_{i,j} X_{i,j}^{(b)} Z_{i,j+1}^{(a)}Z_{i,j}^{(a)}  Z_{i-1,j}^{(a)}
\end{eqnarray}
which describes the the cluster model on the lattice shown in Fig.~\ref{fig:tennet}~(left), and is isomorphic to the honeycomb lattice.
We remark here that the honeycomb lattice cluster model may even be easier to realize practically than the square lattice cluster model, due to a smaller coordination number.
We always consider the SC model on cylinders of circumference $L=2^{l}-1$ along $\vec{e}_1$.

Our second model is the Fibonacci cluster (FC) model, given by the Hamiltonian
\begin{eqnarray}
H_{FC} &=& -\sum_{i,j} X_{i,j}^{(a)}Z_{i,j-1}^{(b)} Z_{i-1,j}^{(b)}Z_{i,j}^{(b)}  Z_{i+1,j}^{(b)} \nonumber  \\
&&-\sum_{i,j} X_{i,j}^{(b)} Z_{i,j+1}^{(a)} Z_{i-1,j}^{(a)}Z_{i,j}^{(a)}   Z_{i+1,j}^{(a)}
\end{eqnarray}
which describes the cluster model on the lattice shown in Fig.~\ref{fig:tennet}~(right).
We always consider this model on cylinders of circumference $L=2^l$ along $\vec{e}_1$.

Let us briefly discuss the symmetries.
First, we define the vector $\mathbf{q}^\alpha (j)$ such that $(\mathbf{q}^\alpha (j))_i = q_{(i,j,\alpha)}$, which has the interpretation of being the \emph{state} of the cellular automaton $\alpha$ at time $j$.
Then, $S(\{q_s\})$ represents a valid symmetry if $\mathbf{q}^{\alpha}(j)$ is a valid space-time trajectory of the cellular automaton:
 $\mathbf{q}^a(j+1) = f \mathbf{q}^a (j)$ and $\mathbf{q}^b(j-1) = \bar f \mathbf{q}^b(j)$ for all $j$, where $f,\bar f$ are the $\mathbb{F}_2$-linear \emph{evolution} operators, defined for the SC acting on a state $\mathbf{q}$ as
\begin{eqnarray}
(f_{SC} \mathbf{q})_i &=& q_i + q_{i-1}; \hspace{0.5cm}
(\bar{f}_{SC} \mathbf{q})_i = q_i + q_{i+1}
\label{eq:fscdef}
\end{eqnarray}
and for the FC as
\begin{eqnarray}
(f_{FC} \mathbf{q})_i &=& (\bar{f}_{FC} \mathbf{q})_i =  q_{i-1} +q_i +  q_{i+1}
\end{eqnarray}
recall that $q_i\in\mathbb{F}_2$, and all addition is modulo $2$.
These rules lead to self-similar fractal structures~\cite{Yoshida2013-of}.  
For example, $f_{SC}$ leads to the Sierpinski gasket at large scales, hence its name.
For the sizes we have chosen, the total symmetry group is simply $(\mathbb{Z}_2\times\mathbb{Z}_2)^{k(L)}$, where $k(L)=L-1$ for the SC, and $k(L)=L$ for the FC. 
The crucial difference between the SC and FC here is that on the specified cylinders, $f_{FC}$ corresponds to a reversible cellular automaton, while $f_{SC}$ does not.
Nevertheless, $f_{FC}$ is  \emph{effectively reversible} when restricted to only even ($\sum_i q_i=0$) states.
As we shall see, in the quantum computation this translates to the identity gate only being realized on the even parity subspace.
We may therefore define the inverse evolution $f^{-1}$, such that $f^{-1} f \mathbf{q} = \mathbf{q}$ for all $\mathbf{q}$ for the FC, while for only even $\mathbf{q}$ for the SC.  
These inverses are discussed in detail in Appendix~\ref{app:finverse}.

These models lie within their own nontrivial SPT phases, protected by the full set of fractal symmetries~\cite{Devakul2018-ru,Kubica2018-dp}.
Next, we demonstrate a scheme for universal MBQC using the unperturbed cluster state, which can then be generalized to elsewhere in the phase.

\section{Measurement based quantum computing with the cluster state}\label{sec:clusstatecomp}
\begin{figure}[t]
\begin{overpic}[
width=0.45\textwidth,tics=10]{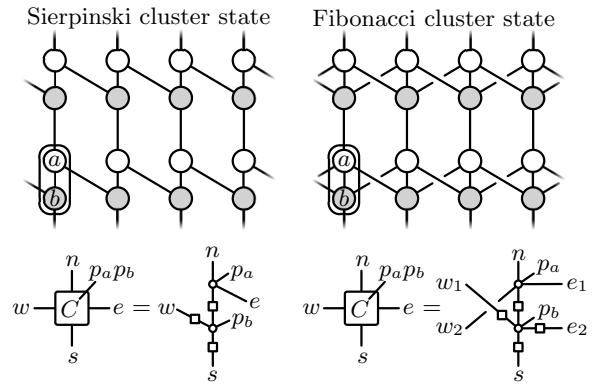}
    \put(26,58){\makebox(0,0){\small Sierpinski cluster state}}
    \put(73,58){\makebox(0,0){\small Fibonacci cluster state}}
    
    \put(13,10){\makebox(0,0){\small $C$}}
    \put(10.5,34.5){\makebox(0,0){\small $a$}}
    \put(10.5,28.4){\makebox(0,0){\small $b$}}
    \put(5,10){\makebox(0,0){\small $w$}}
    \put(13.5,2){\makebox(0,0){\small $s$}}
    \put(21,10){\makebox(0,0){\small $e$}}
    \put(13.5,18){\makebox(0,0){\small $n$}}
    \put(20,16){\makebox(0,0){\small $p_ap_b$}}
    \put(24.2,10){\makebox(0,0){ $=$}}
    \put(36.5,-1){\makebox(0,0){\small $s$}}
    \put(29,10){\makebox(0,0){\small $w$}}
    \put(41.5,8.0){\makebox(0,0){\small $p_b$}}
    \put(43.5,11.0){\makebox(0,0){\small $e$}}
    \put(41.5,16){\makebox(0,0){\small $p_a$}}
    \put(36.5,19.5){\makebox(0,0){\small $n$}}
    
    \put(60.8,10){\makebox(0,0){\small $C$}}
    \put(58.2,34.5){\makebox(0,0){\small $a$}}
    \put(58.2,28.4){\makebox(0,0){\small $b$}}
    \put(52.8,10){\makebox(0,0){\small $w$}}
    \put(61.3,2){\makebox(0,0){\small $s$}}
    \put(68.8,10){\makebox(0,0){\small $e$}}
    \put(61.3,18){\makebox(0,0){\small $n$}}
    \put(67.8,16){\makebox(0,0){\small $p_ap_b$}}
    \put(72,10){\makebox(0,0){ $=$}}
    \put(75,14){\makebox(0,0){ $w_1$}}
    \put(75,6.7){\makebox(0,0){ $w_2$}}
    \put(96,14){\makebox(0,0){ $e_1$}}
    \put(96,6.7){\makebox(0,0){ $e_2$}}
    \put(92,9.5){\makebox(0,0){\small $p_b$}}
    \put(92,17){\makebox(0,0){\small $p_a$}}
    \put(87,-1){\makebox(0,0){\small $s$}}
    \put(87,19){\makebox(0,0){\small $n$}}
\end{overpic}
\caption{The lattices on which the SC (top left) and FC (top right) are simple cluster models (Eq.~\eqref{eq:clusham}).  
In our tensor network description, we group the $a$ and $b$ sites as shown into one tensor $\mathcal{C}[p]=(C_{news}[p])$, indexed by the internal virtual indices for each compass direction ($n$, $e$, $w$, and $s$) and $4$-dimensional physical index $p = p_a p_b$.
For the SC, all virtual indices have dimension $2$, while for the FC, $w = w_1 w_2$ and $e = e_1 e_2$ are $4$-dimensional indices.
$\mathcal{C}^{SC}[p]$ ($\mathcal{C}^{FC}[p]$) is defined according to the tensor network diagrams in the bottom left (right). 
Here, small circles represent a scaled $\delta$ tensor which is $\frac{1}{\sqrt{2}}$ if all indices are equal in the computational ($Z$) basis and $0$ otherwise, and the small squares represent $2\times 2$ Hadamard gates.
}\label{fig:tennet}
\end{figure}
First, we remark that the universality of MBQC with the cluster state is not surprising~\cite{Van_den_Nest2006-lp}.
Following the scheme of Ref.~\onlinecite{Raussendorf2003-vg} it is always possible, via measurements in the $Z$ basis, to effectively isolate 1D chains --- MBQC then follows in a similar manner as for the square lattice.
However, this scheme fails far away from the cluster state fixed point.  
In this section we present a different scheme for universal MBQC, inspired by Ref.~\onlinecite{Raussendorf2018-nh}, which adapts more straightforwardly to elsewhere in the fractal SPT phase.

The computational scheme goes as follows. 
The cluster state is prepared on a long cylinder with circumference $L$ along $\vec{e}_1$ and length $NL$, for large $N$, along $\vec{e}_2$.  
The direction $\vec{e}_2$ is interpreted as the time direction of the quantum circuit.
All physical spins in each $L\times L$ block are measured, which induces an application of some quantum gate to some number (upper bounded by $L$) of logical qubits in the virtual (computational) space, up to byproduct operators unavoidable in MBQC~\cite{Raussendorf2001-xm,Raussendorf2003-vg}. 
The precise gate depends on the basis in which the measurements are performed.
At the end, the state of the final row of unmeasured $b$ sites contains the full information of the output state of the circuit.

We first introduce a tensor network representation of the FC and SC states.  
These states are described exactly by the translationally invariant tensor networks with tensors $\mathcal{C}[p] = (C_{news}[p])$, defined in Fig.~\ref{fig:tennet}~(bottom).
We use the notation $\mathcal{C}[|p\rangle]$ to denote contraction of the physical index with the state vector $|p\rangle$.
The tensors $\mathcal{C}[p]$ obey the following \emph{cluster-like} symmetries
\begin{eqnarray}
\mathcal{C}^{SC}[|p\rangle] &=& X_n X_e X_s \mathcal{C}^{SC}[X_{p_a} |p\rangle] = X_w X_s \mathcal{C}^{SC}[|p\rangle]\nonumber\\
&=& Z_s Z_w Z_n \mathcal{C}^{SC}[X_{p_b}|p\rangle] = Z_e Z_n \mathcal{C}^{SC}[|p\rangle]\nonumber\\
\mathcal{C}^{FC}[|p\rangle] &=& X_n X_{e_1} X_{w_2} X_s \mathcal{C}^{FC}[X_{p_a} |p\rangle] = X_{e_2} X_s \mathcal{C}^{FC}[|p\rangle]\nonumber\\
 &=&  X_{w_1} X_s \mathcal{C}^{FC}[|p\rangle] = 
  Z_s Z_{e_2} Z_{w_1} Z_n \mathcal{C}^{FC}[X_{p_b} |p\rangle]\nonumber\\ &=& Z_{w_2} Z_n \mathcal{C}^{FC}[|p\rangle] = 
  Z_{e_1} Z_n \mathcal{C}^{FC}[|p\rangle]
  \label{eq:clusterlike}
\end{eqnarray}
for the SC or FC, and the following \emph{cluster} symmetries (for both SC and FC)
\begin{eqnarray}
\mathcal{C}[|p\rangle] &=&   X_s \mathcal{C}[Z_{p_b}|p\rangle] = Z_n \mathcal{C}[Z_{p_a} |p\rangle]\label{eq:clustersym}
\end{eqnarray}
where $X_n$ is the $X$ Pauli matrix operating on the $n$ leg of $\mathcal{C}$, and so on.
Together, all these symmetries are sufficient to fully specify $\mathcal{C}$.  
The cluster-like symmetries will be shown to hold anywhere within the phase, while the cluster symmetries are only true at the cluster state fixed point~\cite{Raussendorf2018-nh}.

We now take the system on a cylinder of circumference $L$.
Consider the transfer matrix $\mathcal{T}[\mathbf{p}] = (T[\mathbf{p}]_{\mathbf{s},\mathbf{n}})$ obtained when the state of all the spins $i$ along a row has been fixed (by measurement) to $\mathbf{p} = (|p_i\rangle )$,
\begin{center}
\makebox(0,40){
\hspace{1.5cm}
\begin{overpic}[
width=0.30\textwidth,tics=10]{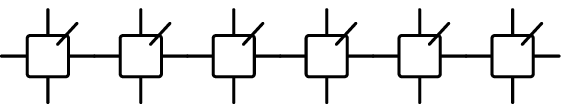}
\put(-25,10){\makebox(0,0){$T[\mathbf{p}]_{\mathbf{s},\mathbf{n}} = $}}
\put(9,19){\makebox(0,0){$n_0$}}
\put(26,19){\makebox(0,0){$n_1$}}
\put(42.0,19){\makebox(0,0){$n_2$}}
\put(58.0,19){\makebox(0,0){$n_3$}}
\put(75.5,19){\makebox(0,0){$n_4$}}
\put(92.0,19){\makebox(0,0){$n_5$}}
\put(9,-2){\makebox(0,0){$s_0$}}
\put(26,-2){\makebox(0,0){$s_1$}}
\put(42.0,-2){\makebox(0,0){$s_2$}}
\put(58.0,-2){\makebox(0,0){$s_3$}}
\put(75.5,-2){\makebox(0,0){$s_4$}}
\put(92.0,-2){\makebox(0,0){$s_5$}}

\put(19,16){\makebox(0,0){\scriptsize $|p_0\rangle$}}
\put(35,16){\makebox(0,0){\scriptsize $|p_1\rangle$}}
\put(51.5,16){\makebox(0,0){\scriptsize $|p_2\rangle$}}
\put(68,16){\makebox(0,0){\scriptsize $|p_3\rangle$}}
\put(85,16){\makebox(0,0){\scriptsize $|p_4\rangle$}}
\put(101.5,16){\makebox(0,0){\scriptsize $|p_5\rangle$}}
\put(-4.5,9.1){\makebox(0,0){\dots}}
\put(105,9.1){\makebox(0,0){\dots}}
\end{overpic}}
\end{center}
where all internal $e$ and $w$ indices have been summed over, and $\mathbf{s}=(s_i)$, $\mathbf{n}=(n_i)$, are the remaining virtual indices, which are combined to form the indices of the matrix $\mathcal{T}[\mathbf{p}]$.

First, consider $\mathcal{T}_0$ where all physical spins have been fixed to $|p_i\rangle = |+\rangle_a|+\rangle_b$ with $|\pm\rangle \equiv (|0\rangle\pm|1\rangle)/\sqrt{2}$.
Let $Z(\mathbf{v}) = \prod_{i=0}^{L-1} Z_i^{v_i}$ be an $L$-qubit Pauli $Z$ operator acting on the virtual subspace, and similarly for $X(\mathbf{v})$, where $v_i\in\{0,1\}$.
Then, Eq.~\eqref{eq:clusterlike} implies that $\mathcal{T}_0$ has the symmetry
\begin{eqnarray}
\mathcal{T}_0 = X(f\mathbf{v}) \mathcal{T}_0 X(\mathbf{v}) = Z(\mathbf{v}) \mathcal{T}_0 Z(\bar{f} \mathbf{v})\label{eq:t0syms}
\end{eqnarray}
for arbitrary vectors $\mathbf{v}$.  These completely specify $\mathcal{T}_0$, which therefore enacts the same cellular automaton, $f$ and $\bar f$, as that of the protecting symmetry.

Now, consider making measurements on all physical spins along this row in the $XY$ plane, such that 
\begin{equation}
|p_i\rangle = \frac{1}{2} \left(|0\rangle_a + (-1)^{\eta_i^{a}} e^{i \delta_i^{a}} |1\rangle_a\right)\left(|0\rangle_b + (-1)^{\eta_i^{b}} e^{i \delta_i^{b}} |1\rangle_b\right) ,\label{eq:pidef}
\end{equation}
 where $\delta_i^{\alpha}$ is the angle in the $XY$ plane of the measurement on the $\alpha$ spin at site $i$ (which we have full control over) and $\eta_i^{\alpha}\in\{0,1\}$ is the measurement result (which we do not have control over).  
 Then, 
\begin{equation}
\mathcal{T}[\mathbf{p}] = \left[\prod_i X_i^{\eta_i^b} e^{i \delta_i^{b} X_i}\right] \mathcal{T}_0  \left[\prod_i Z_i^{\eta_i^a}e^{i \delta_i^{a}Z_i}\right]\label{eq:tp}
\end{equation}

Now, we may begin to discuss computational operations.
One computational step consists of performing measurements on an $L\times L$ block of the cylinder.
This is represented by the matrix
$\mathrm{T} = \prod_{j=0}^{L-1} \mathcal{T}[\mathbf{p}_{j}]$ where $(\mathbf{p}_j)_i$ is the measured state of the $i$th spin in the $j$th row of this block, which we again parameterize by $\delta_{ij}^\alpha$ and $\eta_{ij}^\alpha$, as in Eq.~\eqref{eq:pidef}.

Let us first consider measuring all physical spins in the $X$ basis (all $\delta_{ij}^\alpha=0$), which leads to the realization of the identity gate.
Using Eqs.~\eqref{eq:t0syms} and~\eqref{eq:tp}, we may show that
\begin{eqnarray}
\mathrm{T}_\text{iden} = U_{\Sigma}(\{\eta^\alpha_{ij}\}) \mathcal{I}\label{eq:tiden}
\end{eqnarray}
where $U_\Sigma$, a product of Pauli operators, is the byproduct operator and $\mathcal{I} = \mathcal{T}_0^{L}$.
For the FC, we have $\mathcal{I}^{FC}=\mathds{1}$, the identity gate.
However, for the SC, this is not the case.
Let us define $\overline{X}=\prod_i X_i$ and similarly $\overline{Z}$.  Then, $\mathcal{I}^{SC}= P_{e}+\overline X P_{o}$, where $P_{e(o)}=(\mathds{1}+(-)\overline Z)/2$ is the projector on to the even (odd) subspace.
Thus, $\mathcal{I}^{SC} = \mathds{1}$ when acting on an even state, and $\mathcal{I}^{SC}=\overline{X}$ when acting on an odd state (which then turns it even).


The byproduct operator for the FC is given, up to an overall sign, by
\begin{eqnarray}
U_\Sigma^{FC} &=&  \prod_{j=0}^{L-1} X\left(f^{j}_{FC} \bm{\eta}_{L-1-j}^{b}\right)
\prod_{j=1}^{L-1}Z\left(\bar{f}^{-j}_{FC} \bm{\eta}_{L-j}^{a}\right)
\end{eqnarray}
where $(\bm{\eta}^\alpha_j)_i = \eta^\alpha_{ij}$.  For the SC, this is
\begin{eqnarray}
U_\Sigma^{SC} &=&  \overline{X}^{\sigma(\bm{\eta}_{L-1}^b)}\overline{Z}^{\sigma(\bm{\eta}_{0}^a)} \times \nonumber\\
 &&  \prod_{j=0}^{L-1} X\left( f^{j}_{SC} \widetilde{\bm{\eta}}_{L-1-j}^{b}\right)
  \prod_{j=1}^{L-1}Z\left( \bar{f}^{-j}_{SC} \widetilde{\bm{\eta}}_{L-j}^{a}\right)
\end{eqnarray}
where $\sigma(\bm{\eta}) = \sum \eta_i$ (mod $2$), and $\widetilde{\bm{\eta}} \equiv \mathbf{1}\sigma(\bm{\eta})+\bm{\eta}$ is guaranteed to be even.

If we measure a single physical spin at an angle $\theta$ in the XY plane (setting $\delta_{i_0,j_0}^a$ or $\delta_{i_0,j_0}^b$ to $\theta$), we get
\begin{eqnarray}
    \mathrm{T}_{a,\theta} = U_{\Sigma}(\{\eta^\alpha_{ij}\}) \mathcal{I} e^{\pm i \theta Z(\bar{f}^{j_0}\bm{i_0})}\label{eq:ta}
\end{eqnarray}
for an $a$ spin, or
\begin{eqnarray}
    \mathrm{T}_{b,\theta} = U_{\Sigma}(\{\eta^\alpha_{ij}\}) e^{\pm i \theta X(f^{L-1-j_0}\bm{i_0})} \mathcal{I} \label{eq:tb}
\end{eqnarray}
for a $b$ spin, where $\bm{i_0}$ is a vector with zeros everywhere except at $i_0$.
These may therefore act as single or multi-qubit rotations.
The $\pm$ sign in the exponent arises from commutation with potential byproduct operators, which can be corrected for by choosing $\theta\rightarrow \pm \theta$ if all measurements with $j<j_0$ have been made.

The full computation begins with an initialization of the state (which can be done by measuring the $a$ sites of the first row of the first block in the $Z$ basis).
Each $L\times L$ block then implements a unitary gate according to Eq.~\eqref{eq:ta} or~\eqref{eq:tb}.  
Finally, at the end of the computation, the last row of unmeasured $b$ sites contains the computation result, up to byproduct operators and an overall Hadamard transformation (this Hadamard transformation can be avoided by adding a final row of $a$ sites which then contains the output state after the final block is measured).

\section{Universality} 
To prove universality, let us consider the elementary gates, which are obtained by setting a single $\delta_{ij}^{\alpha}=\theta$, while keeping the rest 0.
We first discuss the FC.
Using non-zero $\delta_{k,0}^a$ or $\delta_{k,L-1}^b$ results in an arbitrary single qubit rotation about the $Z$ or $X$ axis, $e^{i\theta Z_k}$ or $e^{i\theta X_k}$, from which all single-qubit unitary gates can be obtained (assuming the $\pm$ sign from the byproduct operator has been corrected for).  
Next, using  $\delta_{k,1}^a$ results in the unitary $e^{i\theta Z_{k-1}Z_kZ_{k+1}}$. 
Note that it is \emph{not} possible to directly perform any two-qubit entangling gates.
One of many ways to obtain universality is to use only the even $L/2$ qubits as our \emph{logical} qubits, $\ell^Z_k = Z_{2k}$ for $k=0\dots L/2-1$.
Then, after initialization in the $Z$ basis, every odd qubit can be rotated into the $Z=1$ state.
On the logical qubits, arbitrary single qubit unitaries and $e^{i\theta\ell^Z_k \ell^Z_{k+1}}$ are possible, which constitutes a universal set on $L/2$ logical qubits.

Some subtleties arise in the case of the SC (as to be expected).
Without loss of generality, let us work only with the even parity computational states ($\overline{Z}=1$).
The only time a state is not purely even is potentially right after initialization or after a $\mathrm{T}_{b,\theta}$ operation (Eq.~\eqref{eq:tb}), but we can always follow up with a $\mathrm{T}_\text{iden}$ to restore the total parity (the byproduct operator $U^{SC}_\Sigma$ may change the parity of the state, but recall that we propagate all byproduct operators to the end of the computation where they are corrected for post-measurement).
Replacing $\mathrm{T}_{b,\theta}\rightarrow \mathrm{T}_\text{iden}\mathrm{T}_{b,\theta}$ has the effect that when $\delta_{k,L-1}^b=\theta$, the gate $e^{i\theta X_k}\rightarrow e^{i\theta \overline{X} X_k}$ no longer corresponds to a single qubit unitary.
Universality on $(L-1)/2$ \emph{logical} qubits may be established by defining $\ell^Z_{k} = Z_{2k} Z_{2k+1}$ for $k=0\dots (L-3)/2$.
Using $\delta_{2k,1}^{a}$, one obtains $e^{i\theta \ell^Z_{k}}$, and using $\delta_{2k,L-1}^{b}$, one obtains $e^{i\theta \ell^X_k}$, which are universal for single qubits (where $\ell^X_k\equiv \overline X X_{2k}$).
Then, using $\delta_{2k,3}^{a}$, we get $e^{i\theta \ell^Z_{k} \ell^{Z}_{k+1}}$ (for all $k$ except $k=(L-3)/2$).
These constitute a universal set of gates on $(L-1)/2$ logical qubits.
Furthermore, many other gates are easily realized in single measurement steps --- for example, two logical qubits separated by a large power of $2$ may be entangled by a two-qubit gate in a single step. This feature may have useful practical applications.

\section{Away from the cluster fixed point}
We have carefully set up our MBQC scheme such that it may be easily extended away from the cluster state fixed point, provided the full set of fractal symmetries are respected.
From here on, the proof for universality throughout the fractal SPT phase follows that of the square lattice cluster model in Ref.~\onlinecite{Raussendorf2018-nh} without issue.
We briefly outline the proof here.

The set of fractal symmetries pose a strict constraint on the possible allowed perturbations.
Any state in the fractal SPT phase, $|\psi\rangle$, may be connected to the cluster state $|\mathcal{C}\rangle$ via a finite depth symmetry-respecting local unitary circuit, $|\psi\rangle = U|\mathcal{C}\rangle$.
Expanding $U$ in the Pauli basis of $L\times NL$ spins, $U = \sum  c\,  X(\cdot)  Z(\cdot)$, the only symmetry respecting $ Z(\cdot)$ terms must be products of $\prod_{s^\prime \in\Gamma(s)}Z_{s^\prime}$ (a proof of this claim is in Appendix~\ref{app:zpert}).  
Making use of a property of $|\mathcal{C}\rangle$, $\prod_{s^\prime \in\Gamma(s)}Z_{s^\prime}|\mathcal{C}\rangle=X_s|\mathcal{C}\rangle$, leads to the following fact~\cite{Raussendorf2018-nh}:
Anywhere in the fractal SPT phase the tensor network state is described by tensors $\mathcal{A}^{ij}[|p\rangle]$ (where $ij$ labels the unit cell) which have the property that for $|p\rangle = |\pm_{a} \pm_{b}\rangle$ in the \emph{symmetry-protected} $X$ basis,
\begin{equation}
    \mathcal{A}^{ij}[|\pm_a \pm_b\rangle] = \mathcal{B}^{ij}[|\pm_a \pm_b\rangle] \otimes \mathcal{C}[|\pm_a \pm_b\rangle]
\end{equation}
factors into a non-universal \emph{junk} part, $\mathcal{B}$, and the universal cluster part $\mathcal{C}$ from earlier.
Crucially, the symmetry operators $X_a$ and $X_b$ act trivially on $\mathcal{B}$.  
$\mathcal{A}^{ij}$ therefore obeys all our cluster-like symmetries, Eq.~\eqref{eq:clusterlike} (note that we never used the cluster symmetries from Eq.~\eqref{eq:clustersym} in any of our arguments).
This alone is enough to prove that the identity gate $\mathrm{T}_\text{iden}$ can be realized exactly, as in Ref.~\onlinecite{Else2012-ie}, and this state therefore acts as a \emph{quantum wire} on $k(L)$ qubits.
To perform non-identity gates, the \emph{oblivious wire} is used to turn quantum wire into computation~\cite{Raussendorf2017-gb}.
In this procedure, a unitary evolution is maintained only to first order in the angle $\delta$ away from the $X$ axis.
Thus, a unitary rotation by angle $\theta$ is accomplished by $n$ repeated rotations of a small angle $\theta/n$.
The measurement and initialization procedure should also be modified accordingly~\cite{Raussendorf2017-gb}.
The computational scheme presented in previous sections generalize in a straightforward manner to this type of procedure away from the fixed point.


\section{Why rigid subsystem symmetries?}
The schemes considered here and in Ref.~\onlinecite{Raussendorf2018-nh} are qualitatively different to previous approaches to universal MBQC in two dimensions.
Previous approaches~\cite{Darmawan2012-wg,Huang2016-lo,Wei2017-gy} for performing 2D MBQC in the presence of perturbations essentially rely upon distilling an almost exact cluster or valence-bond state via measurement and then using further measurements to decouple effective quantum wires and perform entangling gates between them. 
Here we instead consider resource states that reduce, on a long cylinder, to quantum wires for a number of qubits that grows with the radius, without the requirement that regions of qubits are measured in the $Z$ basis to decouple quantum wires. 
This allows results developed for 1D SPT quantum wires to be applied~\cite{Else2012-ie,Miller2015-gl,Stephen2017-cp,Raussendorf2017-gb,Raussendorf2018-nh}. 
In this section we demonstrate the importance of rigid  (line~\cite{You2018-em} or fractal~\cite{Devakul2018-ru}) subsystem symmetry, as opposed to global or $(\text{D}-1)$-form (deformable line) symmetries~\cite{Gaiotto2015-cj}.

Suppose we have a unique short-range entangled ground state $\ket{\psi}$ of a gapped local Hamiltonian that is symmetric under a $(\text{D}-1)$-form symmetry $U_g^\lambda$, for $g\in G$ and $\lambda$ a closed path on the lattice. 
 Applying $U_g^\gamma$ to $\ket{\psi}$ along an open path $\gamma$, with domain $[0,1]$, creates excitations in the neighborhood of its end points (possibly located at lattice boundaries). These exciations can be locally annihilated by some operators $V_g^{\gamma_0},\, V_g^{\gamma_1},\,$ with support size on the order of the correlation length, i.e. 
 \begin{align}
 (V_g^{\gamma_0} \otimes V_g^{\gamma_1})  
 U_g^\gamma \ket{\psi} = \ket{\psi} 
 \, .
 \end{align}
  Unlike in 1D, where these end point operators may form a projective representation~\cite{Gu2009-ly,Chen2010-sm,Chen2011-ss,Chen2012-oa,Pollmann2012-lv,Schuch2011-gq,Chen2013-gq}, in 2D or higher we can consider a disjoint path $\gamma'$ sharing an end point with $\gamma$, without loss of generality assume $\gamma_0=\gamma'_0$, and the other end points separated by a distance much larger than the correlation length. 
 Then $[V_g^{\gamma_0},V_h^{\gamma'_0}]=0$ since $[U_g^\gamma,\, U_h^{\gamma'}]=0$ and $[ V_g^{\gamma_1},\,  V_h^{\gamma'_1}]=0$ as they are pairs of operators with disjoint support. 
Furthermore, since the symmetry is deformable the endpoint operator should only depend on the endpoint location $\gamma_0'=\gamma_0$. This implies $[V_g^{\gamma_0},V_h^{\gamma_0}]=0$ and hence $V_g$ cannot form a nontrivial projective representation as all the matrices commute. 
 Therefore, when the state is viewed as a 1D SPT on a long cylinder, with respect to the $(\text{D}-1)$-form  symmetries running along the cylinder, it must lie in the trivial phase and generically will not be useful as a quantum wire. 

If one additionally considers a global symmetry (such as the cluster state in the Appendix of Ref.~\onlinecite{Devakul2018-ru}), the boundary operators for the $(\text{D}-1)$-form symmetry do not necessarily commute with the boundary operators for the global symmetry. This can lead to nontrivial projective representations at the end of a long cylinder and hence a nontrivial 1D SPT phase under the combined global and $(\text{D}-1)$-form symmetries along the cylinder. 
However, the 1D SPT phases produced in this way can only support a constant stable edge degeneracy as the radius of the cylinder increases, and hence can only wire a constant number of qubits. 
Consequently, for schemes such as the one considered in this paper and in Ref.~\onlinecite{Raussendorf2018-nh} global and $(\text{D}-1)$-form symmetries do not suffice and rigid subsystem symmetries are necessary for robust MBQC on a number of qubits growing with $L$.

Note that mere existence of a nontrivial subsystem SPT phase alone does not imply that a universal set of logical gates are possible via single spin measurements --- this is a property of the underlying cellular automaton.
For example, consider the 2D phase consisting of decoupled 1D SPT chains oriented vertically (a ``weak subsystem SPT''~\cite{You2018-em}).  
On a cylinder this phase serves as a quantum wire for a number of qubits growing with $L$, but entangling gates between qubits from different chains cannot be accomplished using only single qubit measurements.

\section{Conclusion}
We have shown that 2D cluster models with fractal symmetries, exemplified here by the SC and FC, may serve as resources for universal measurement-based quantum computation.
Furthermore, this is a property of the entire fractal SPT phase, not just the cluster state fixed point.
Despite the fractal structure of the symmetries, we reiterate that the underlying models are simple cluster models on regular lattices.  

Further questions involve other types of symmetries.  
The square lattice cluster model in the proof of Ref.~\onlinecite{Raussendorf2018-nh} is protected by rigid \emph{subsystem symmetries}~\cite{You2018-em}.
These are fundamentally different from (seemingly similar) $(\text{D}-1)$-form symmetries~\cite{Gaiotto2015-cj}, as we have shown.
A particular 2D cluster model possessing global and $(\text{D}-1)$-form symmetries on a cylinder can only wire a single qubit ---
is there a scheme by which such a model is useful for universal quantum computation beyond small perturbations of the cluster model fixed point?

Another interesting question is whether there exists a rigid subsystem SPT MBQC scheme for a 3D model where the boundary qubits are topologically protected. This may allow rigid subsystem SPT MBQC to persist to nonzero temperatures~\cite{PhysRevA.71.062313,PhysRevA.96.022306}.

\section*{Acknowledgements}
DW thanks Nick Bultinck for his hospitality while visiting Princeton. 
\paragraph*{Funding information}
TD is funded by the DOE SciDAC program, FWP 100368 DE-AC02-76SF00515

\begin{appendix}

\section{Reversing the cellular automata} \label{app:finverse}
Here, we discuss reversing the evolution of the cellular automata $f_{FC}$ and $f_{SC}$ on rings of circumference $L=2^{l}$ and $2^{l}-1$, respectively.

It is helpful to use a polynomial representation to describe these linear cellular automata. 
Let us define the Laurent polynomial $q(x)$ over $\mathbb{F}_2$ corresponding to the state $\mathbf{q}$ as $q(x) = \sum_{i=0}^{L-1} q_i x^i$.  
Periodicity is enforced by setting $x^{L}=1$.  
In this language, the evolution of the cellular automaton is encoded in a single polynomial $f(x)$, such that if $q_t(x)$ describes the state at time $t$, then the state at the next time is described by $q_{t+1}(x) = f(x) q_t(x)$.  
The evolution $\bar{f}$ in the main text is obtained by $\bar f(x) = f(x^{-1})$.

For the FC, $f_{FC}(x) = x^{-1}+1+x$.  
In particular, suppose we apply the evolution $2^{l-1}=L/2$ times, we have that $f_{FC}^{L/2}(x) = x^{-L/2}+1+x^{L/2}=1$,
where we have used the fact that $x^{-L/2} = x^{L/2}$, and that the binomial coefficient ${L/2 \choose k}$ is $0$ (mod $2$) for all $k$ except $k=0,L/2$.
Thus, $f_{FC}^{L/2}(x)=1$, which therefore implies that $f_{FC}$ is reversible, and the inverse evolution is given by $f_{FC}^{-1}(x) \equiv f_{FC}^{L/2-1}(x)$.
Starting with an arbitrary state $q_t(x)$, this guarantees that it will have a cycle with period $2^{l-1}$, $q_{t+L/2}(x) = q_t(x)$.
In the main text, for simplicity, we took the fundamental computational step to be an $L\times L$ block --- for the FC this can be reduced to $L\times (L/2)$.

For the SC, $f_{SC}(x)=1+x$.  
Applying it $2^{l}-1=L$ times, we have 
\begin{eqnarray}
f_{SC}^{L} &=& 1+x+x^2+\dots+x^{L-1}+x^{L} \nonumber \\
           &=& x+x^2+\dots+x^{L-1}
\end{eqnarray}
Letting $u(x) = \sum_{i=0}^{L-1} x^i$, we have that $f_{SC}^{L}(x) = 1 + u(x)$, where
$u(x)$ has the property that $x^i u(x) = u(x)$.
Now, consider the action of $f_{SC}^{L}$ on an \emph{even} state, which we define as a state with $q(0) = \sum_{i}q_i=0$.
This may be written as 
\begin{eqnarray}
f_{SC}^{L}(x) q(x) &=& (1+u(x)) q(x) \nonumber \\
&=& q(x) + \sum_{i=0}^{L-1} q_i x^i u(x) \nonumber \\
&=& q(x) + \left(\sum_{i=0}^{L-1} q_i\right) u(x) \nonumber \\
&=& q(x) + q(0) u(x) = q(x)
\end{eqnarray}
Thus, $f_{SC}^{L}$ acts as identity on any even state.
We may therefore define the inverse $f_{SC}^{-1}(x) \equiv f_{SC}^{L-1}(x)$ which reverses the evolution of $f_{SC}$ when acting on an even state.
This also implies that for any even initial state $q_t(x)$, the SC has a cycle with period $L$, $q_{t+L}(x)=q_t(x)$.

Finally, for completeness, we give the form of the inverse evolution explicitly.
They are given by
\begin{eqnarray}
(f_{SC}^{-1} \mathbf{q})_i &=& \sum_{j=0}^{i-1} z_2^j q_j + \sum_{j=i}^{L-1} z_2^{1+j} q_j \\
(\bar{f}_{SC}^{-1} \mathbf{q})_i &=& (f_{SC}^{-1}\mathbf{q})_i + q_i\\
(f^{-1}_{FC} \mathbf{q})_i &=& (\bar{f}_{FC}^{-1} \mathbf{q})_i =  \sum_{j=0}^{L-1} (1-z_3^{|i-j|-2})q_j
\end{eqnarray}
where $z_m^n=1$ if $n$ is a multiple of $m$, else $0$.

The number of symmetries for the FC and SC models are given by $2^{2 k(L)}$, where $2^{k(L)}$ is the number of distinct cycles of $f$.
For the FC, we saw that for any arbitrary state, $q_{t+L/2}(x)=q_t(x)$, and so $k(L)=L$.
For the SC, we found that only for \emph{even} states, $q_{t+L}(x)=q_t(x)$, while an odd state never returns to itself, and so $k(L)=L-1$.



\section{Restrictions on the form of symmetric $Z$ perturbations}\label{app:zpert}
Here, we prove the claim that any symmetry-respecting operator consisting of only $Z$ Pauli operators, $ Z(\mathbf{v})=\prod_{s} Z^{v_s}$, must be composed of a product of $F_s\equiv \prod_{s^\prime \in \Gamma(s)} Z_{s^\prime}$.
First, consider a $\mathcal{O} = Z(\cdot)$ term purely on the $a$ sublattice, localized within some $L\times L_y$ block.  
The $F$ operators acting on the $a$ sublattice spins are given by
\begin{equation}
    F_{ij}^{SC} = Z_{i,j+1} Z_{i,j} Z_{i-1,j}
\end{equation}
for the SC, and  
\begin{equation}
    F_{ij}^{FC} = Z_{i,j+1} Z_{i-1,j} Z_{i,j} Z_{i+1,j}
\end{equation}
for the FC, where $Z_{i,j}$ acts on the $a$ sublattice spin at the $(i,j)$th unit cell.
Utilizing the fact that $F_{ij}$ consists of only one $Z$ operator on the $(j+1)$th row, and the remaining on the $j$th row, we can use products of $F_{ij}$ to move any $Z$ operator in $\mathcal{O}$ from the bulk of the $L\times L_y$ block to some product of $Z$ operators acting only on the top row.  
Thus, $\mathcal{O}$ within an $L\times L_y$ block can be related to some operator $\mathcal{O}^\prime$ acting only on the top row by applications of $F_{ij}$.  
As $\mathcal{O}$ respects all fractal symmetries, so does $\mathcal{O}^\prime$.

However, there are $2^{k(L)}$ distinct symmetries acting on only the $a$ sublattice on a cylinder of circumference $L$.
By the cellular automaton analogy, knowing how the symmetry acts on one row fully determines its action on all other rows.
For the FC, $k(L)=L$, and so there are $2^L$ distinct  symmetries.  
Hence, on the top row, each of the $2^L$ possible $X(\cdot)$ operators appear in some symmetry.
If $\mathcal{O}^\prime$ (which is a product of $Z$s on the top row) is to commute with all such symmetries, it must be the identity.
For the FC, $k(L)=L-1$.  
The $2^{L-1}$ distinct symmetries acting on the top row are all possible $X(\cdot)$ operators that are tensor products of an even total number of $X$s.  
Therefore, $\mathcal{O}^\prime$ must be either identity or $\prod Z$, a product of $Z$s along the whole top row.
A product of all $Z$s on a row may be eliminated by $F_{ij}$.

Hence, all $\mathcal{O}$ on the $a$ sublattice can be connected to identity by applications of $F_{ij}$ and therefore are composed of $F_{ij}$.  
A similar procedure applies for operators on the $b$ sublattice (now evolving down to the bottom row).

\end{appendix}



\bibliography{Refs}

\end{document}